# Momentum Transfer by Laser Ablation of Irregularly Shaped Space Debris

Duane A. Liedahl, Stephen B. Libby, Alexander Rubenchik

Lawrence Livermore National Laboratory, 7000 East Ave, Livermore, CA 94550, USA

**Abstract.** Proposals for ground-based laser remediation of space debris rely on the creation of appropriately directed ablation-driven impulses to either divert the fragment or drive it into an orbit with a perigee allowing atmospheric capture. For a spherical fragment, the ablation impulse is a function of the orbital parameters and the laser engagement angle. If, however, the target is irregularly shaped and arbitrarily oriented, new impulse effects come into play. Here we present an analysis of some of these effects.

## INTRODUCTION

Of the ~150,000 objects larger than 1 cm diameter in low Earth orbit, the vast majority consists of space debris, remnants of larger manmade objects, a combination of operational debris and fragmentation debris [1]. High-velocity impacts of space payloads with objects with masses as small as a few grams can have potentially devastating effects. Thus the continuing accumulation of space debris in low Earth orbit poses an escalating threat to space assets.

A candidate remediation strategy calls for de-orbiting the debris using a ground-based laser [2]. Among the key components of this concept is the transfer of momentum to debris fragments through surface ablation. The link between laser energy and target momentum change is the mechanical coupling coefficient, by which one embeds the microphysics into a single quantity, $C_m$, defined so that

$$m\Delta v = C_m E_L \qquad (1)$$

where $m$ is the debris mass, and $E_L$ is the laser energy (per impulse) incident on the target. Both theoretical and experimental efforts have provided us with a basic understanding of the relevant microphysics determining the behavior of $C_m$ [4]. To reduce the perigee to the 200 km atmospheric capture altitude, a velocity reduction ~100-200 m s$^{-1}$ must be imposed. The efficiency of the momentum transfer is known to vary with laser intensity and material [3], but a $\Delta v$ of approximately 0.1–1 m s$^{-1}$ for 10 Joules of laser energy on a 1 g target is a fair estimate.

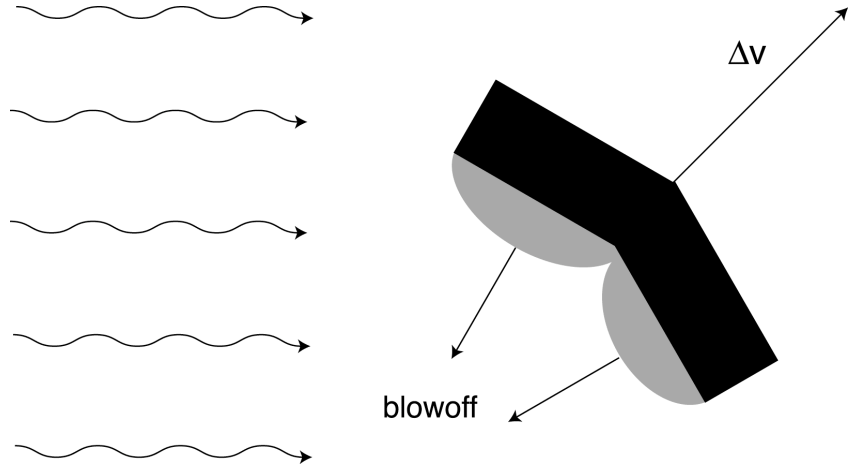

**Figure 1:** Schematic of laser/matter interaction with an irregularly shaped target, in this case, a wedge, shown in black. Plasma blowoff regions, with net velocity vectors parallel to the respective surface normals, are shown in gray. The impulse vector is not aligned with the beam. Arbitrarily shaped targets respond in a variety of ways, depending not only on their shape but also on their orientation with respect to the beam.

Equation (1) is written in scalar form, which implies that the target recoil is precisely parallel to the laser beam. This is, of course, *not* a naïve assumption, but rather an expedient, stemming from both theoretical and experimental studies in which the theory is typically treated as a 1-D phenomenon, and laboratory targets are generally planar. In such cases, targets do, in fact, recoil along the beam, and the numerical estimate given above applies. Similarly, concept studies of debris clearing invoke spherical targets, which, lacking a characteristic vector, also recoil in the beam direction.

If the ablation flow is along the local surface normal, and if the momentum transfer is precisely opposite, then it is easy to imagine a case where the recoil and the beam are misaligned. Figure 1 shows one hypothetical example.

The Orion Project [2] grouped the debris into five "likely" compositional classes: aluminum, steel, Na/K metal, carbon phenolics, and insulation, only a fraction of which possess spherical symmetry, and none of which will consist of planes with surface normals oriented exactly along the beam. Metal fragments, in particular, are likely to comprise a large variety of shapes. We have begun some preliminary studies that relax the assumption of spherical symmetry. The next section provides a discussion of a few basic concepts associated with the impulse response of irregular targets. Following that, we provide an example of the orbital-scale implications for the envisioned de-orbiting scheme.

## Area Matrix Approach

The scalar form of the impulse equation must be replaced by a vector form. Since, in the general case, the impulse vector and the propagation vector of the laser are not parallel, intervention by something resembling a rank two tensor is suggested. Based on the assumption that ablation is parallel to the local normal, and that the impulse is directed exactly opposite to the net ablation vector, we can write

$$m\Delta \vec{v} = -C_m F_L \sum_\alpha A_\alpha \left| \hat{k} \bullet \hat{n}_\alpha \right| \hat{n}_\alpha \tag{2}$$

where the sum is over all illuminated surfaces, each surface with respective area $A_\alpha$, and the laser fluence is given by $\vec{F}_L = F_L \hat{k}$. However, since illumination requires that $\hat{k} \bullet \hat{n} < 0$, we can drop the absolute value bars and the negative sign, which leaves

$$m\Delta \vec{v} = C_m \vec{F}_L \bullet \ddot{G} \tag{3}$$

where $\ddot{G}$ is the dyadic form with an areal dimension, given by

$$\ddot{G} = \sum_\alpha A_\alpha \hat{n}_\alpha \hat{n}_\alpha \tag{4}$$

In the case of a continuously varying normal, this is

$$\ddot{G} = \int dA \, \hat{n}\hat{n} \tag{5}$$

where the integral is performed over the illuminated portion of the surface. We refer to $\ddot{G}$ as the *area matrix*.

## *Example: Sphere*

A spherical fragment with radius $R$ provides a simple example. In this case, the area matrix is

$$\ddot{G} = \int dA \, \hat{r}\hat{r} \tag{6}$$

If we resolve $\hat{r}\hat{r}$ into Cartesian components, and use the standard spherical coordinates, we have

$$\hat{r}\hat{r} = \begin{pmatrix} \sin^2\theta \cos^2\phi & \sin^2\theta \sin\phi \cos\phi & \sin\theta \cos\theta \cos\phi \\ \sin^2\theta \sin\phi \cos\phi & \sin^2\theta \sin^2\phi & \sin\theta \cos\theta \sin\phi \\ \sin\theta \cos\theta \cos\phi & \sin\theta \cos\theta \sin\phi & \cos^2\theta \end{pmatrix} \tag{7}$$

Integration over the irradiated hemisphere shows that the off-diagonal elements vanish, and that the diagonal elements are equal, leaving

$$\ddot{G} = \frac{2\pi}{3} R^2 \, \ddot{I} \tag{8}$$

where $\ddot{I}$ is the identity matrix. Therefore, with $\vec{F} = F_L \hat{k}$, we find that, for a sphere,

$$m\Delta\vec{v} = \frac{2}{3} C_m F_L \pi R^2 \; \hat{k} \qquad (9)$$

As expected, the recoil is parallel to the beam, but note that the impulse is not simply the product of $C_m F_L$ and the projected area. The area matrix accounts for the variation in projected area relative to the beam direction, the direction of the impulse, and the magnitude of the impulse scaled to the laser fluence and the coupling coefficient. One could say in this context that the "effective area" of a sphere is $2\pi R^2/3$.

## *Example: Cylinder*

A more interesting example is provided by a circular cylinder of radius $R$ and height $H$. Assume that the cylindrical axis coincides with the $\hat{z}$-axis. Then, using the standard cylindrical coordinates $\rho$, $z$, and $\phi$, the area matrix is

$$\ddot{G} = \pi R^2 \; \hat{z}\hat{z} + \int dA \; \hat{\rho}\hat{\rho} \qquad (10)$$

$$\hat{\rho}\hat{\rho} = \begin{pmatrix} \cos^2\phi & \sin\phi\cos\phi & 0 \\ \sin\phi\cos\phi & \sin^2\phi & 0 \\ 0 & 0 & 0 \end{pmatrix} \qquad (11)$$

Note that only one cylinder cap is accommodated in this expression for the area matrix, since a unidirectional beam can illuminate only one. After integration, addition of the two terms gives

$$\ddot{G} = \frac{\pi}{2} \begin{pmatrix} RH & 0 & 0 \\ 0 & RH & 0 \\ 0 & 0 & 2R^2 \end{pmatrix} \qquad (12)$$

Figure 2 illustrates, for a particular $R/H$ ratio, the behavior of an irradiated cylinder for a variety of orientations. Both the direction and the magnitude of the recoil depend on the relative orientation. For the cases of illumination parallel or perpendicular to the beam, the recoil is along the beam, albeit with different magnitudes.

For the special case $H = 2R$, the area matrix is simply $\pi R^2 \ddot{I}$; $\Delta\vec{v}$ and $\hat{k}$ are parallel for all incident angles – the cylinder becomes "sphere-like," with an effective cross-section of $\pi R^2$. Note that for this special case, although the geometrical cross-section projected onto the beam direction changes with orientation, the effective cross-section is strictly constant. In other words, recalling Eq. (1), even though the energy incident on the target varies, the momentum change does not, which is a simple example of the inadequacy of the scalar equation. It is easy to show that the same holds true for a cube. Regardless of the orientation, the recoil is along the beam, with magnitude $C_m F_L s^2$, where $s$ is the length of a side. Any object for which the area matrix is a multiple of the identity matrix will share this trait.

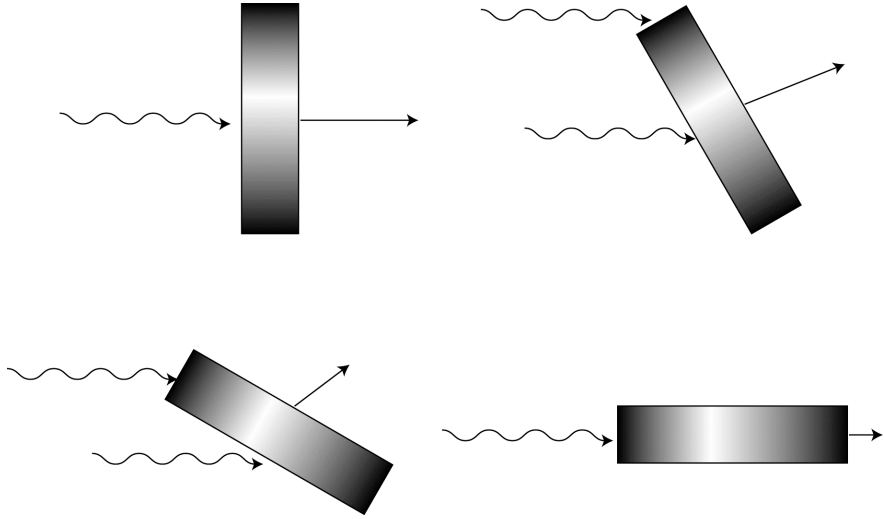

**Figure 2.** Examples of impulse response of a cylinder (side view) for different illumination angles (0°, 30°, 60°, and 90°). The radius of the cylinder is twice the height ($R = 2H$). Directions and magnitudes of the impulse vectors (*straight arrows*) are accurately portrayed, based on Eq. 12. The magnitudes, for this case, show a factor of 4 variation, from the largest (*upper left*) to the smallest (*lower right*).

Another special case, but for arbitrary $H/R$ is a beam that is perpendicular to the cylindrical axis, say $\vec{F}_L = F_L \hat{x}$. The projected area in this case is $2RH$. The result is

$$m\Delta\vec{v} = \frac{\pi}{4} C_m F_L (2RH)\, \hat{x} \tag{13}$$

The factor $\pi/4$ can be compared to the analogous factor of 2/3 for a sphere.

## Target Response to Laser Irradiation: Orbital Scale

The existence of irregularly shaped space debris brings a degree of randomness into the problem of calculating post-engagement orbital modifications: that associated with the distribution of object shapes, and that associated with orientation, including possible tumbling motion. We note that a tumbling non-spherical target may experience a variety of vector impulses during an engagement campaign, if the total duration of the engagement is comparable to or longer than the rotation period of the object. Given the desire to reduce perigee, it is of interest to characterize the range of possible orbital outcomes of laser engagements with non-spherical targets (see Figure 3).

To begin to quantify the concept, we choose the simple case of a plate of mass $m$, in a low elliptical orbit, with eccentricity given by

$$\varepsilon = \left(1 + \frac{2El^2}{G^2 M^2 m^3}\right)^{1/2} \tag{14}$$

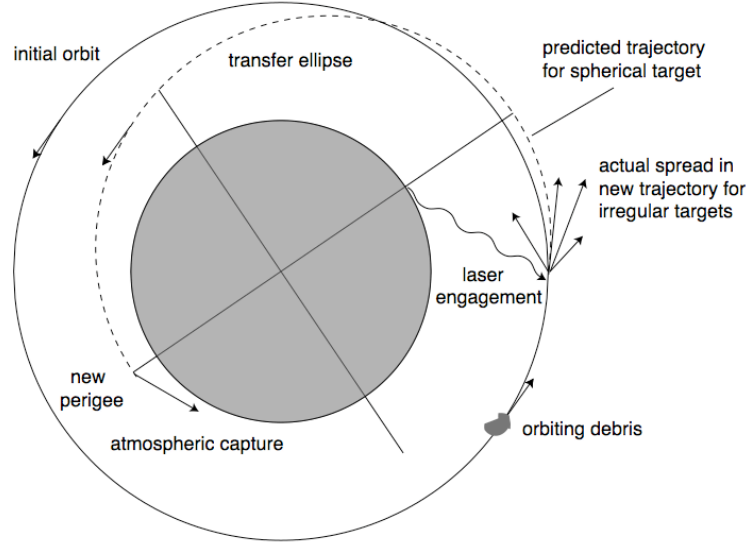

**Figure 3.** Schematic of the de-orbiting concept for space debris in low-Earth orbit (scales exaggerated for clarity). For a given energy deposition, the orbital perturbation on a spherical target is predictable. For non-spherical targets, the perturbation is unknown *a priori*, unless the shape and orientation at engagement are known.

where $E$ is the total orbital energy, $l^2$ is the square of the orbital angular momentum, $G$ is the gravitational constant, and $M$ is Earth's mass. After engagement, a new orbit is determined from changes to $E$ and $l^2$. If the instantaneous distance from Earth's center, orbital velocity, and azimuthal velocity are denoted $r$, $\vec{v}$, and $v_\phi$, respectively, then $\Delta E$ and $\Delta l^2$ are given in terms of the velocity change by

$$\Delta E = m\vec{v} \bullet \Delta\vec{v} + \frac{1}{2} m|\Delta\vec{v}|^2 \qquad (15)$$

$$\Delta l^2 = 2m^2 r^2 v_\phi \Delta\vec{v} \bullet \hat{\phi} + m^2 r^2 \left(\Delta\vec{v} \bullet \hat{\phi}\right)^2 \qquad (16)$$

Note that the second terms in the above equations are much smaller than the first, with relative magnitudes $\Delta v/v \sim 10^{-3}$. For most cases of interest, neglect of the $2^{nd}$-order terms is justifiable – an example for which the $2^{nd}$-order terms dominate is engagement near laser zenith, which would qualify as a waste of laser energy.

The quantity of primary interest is the perigee, which is

$$r_p = \frac{l^2}{GMm^2} \frac{1}{1+\varepsilon} \qquad (17)$$

We calculate the perigee change for a random distribution of plate orientations, and for a representative set of orbital parameters, setting $m = 1$ g for this example. The maximum energy absorbed from the laser is 10 J, which occurs when the plate is face-on to the laser position. The distribution in the perigee change at a fixed orbital angle (not shown) is weakly peaked, with substantial probability at the upper and lower bounds. Thus one can estimate the probability of achieving an undesirable result, i.e., increasing the perigee, by comparing the magnitude of the upper and lower envelopes. It is also worth noting that there is a non-negligible probability of achieving a result that is *more* favorable than for the spherical case. The average perigee change for a plate is approximately 1/3 that found for a sphere, which has implications for the efficiency of targeting campaigns.

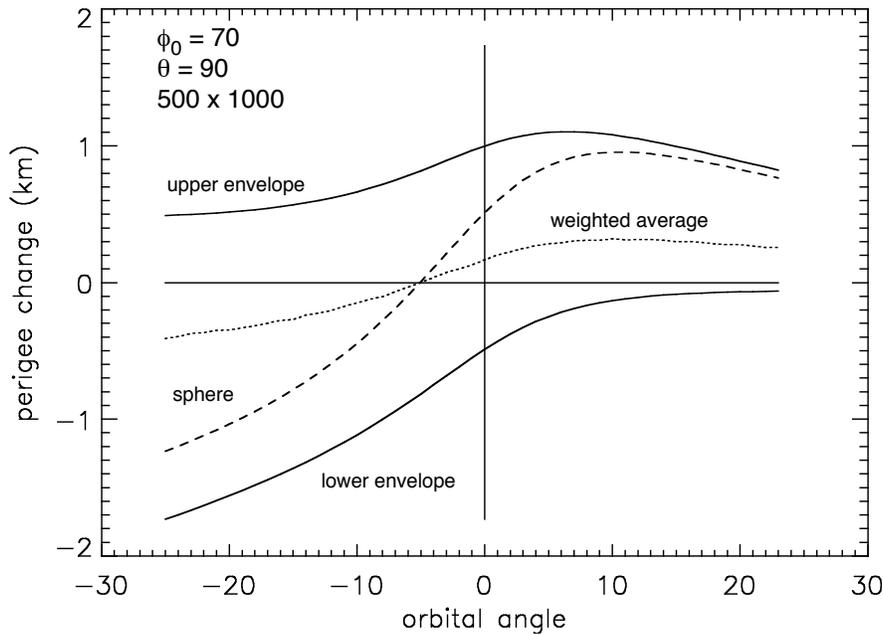

**Figure 4.** Perigee change for a 1 g plate following a nominal 10-Joule-on-target laser interaction, with $C_m$ =10 dyne/W, assuming a random distribution of plate orientations in three dimensions, plotted against orbital angle. Negative angles correspond to upstream positions relative to the laser position at $\phi=0$. Horizontal extent of abscissa maps to horizons. Example orbit is characterized by 500 km perigee, 1000 km apogee, perigee angle ($\phi_0$) 70 degrees downstream of the laser position (descending), orbit intersects laser zenith. Plotted are the best case ("lower envelope"), worst case ("upper envelope"), the weighted average (dotted), and the single-valued result for a spherical target (dashed).

## CONCLUSIONS

Continued definition and refinement of large-scale plans for space debris clearing by laser ablation can benefit from broader characterizations of the behavior of debris fragments subjected to pulsed laser irradiation. This must allow for the reality that most debris fragments are neither spherical nor face-on planes. Here we have presented the most basic considerations required in order to go beyond these simple assumptions. A number of other issues related to shape irregularity need to be addressed.

For example, for an object consisting of a distribution of surface normal vectors, the intensity distribution will vary across the surface ($I \propto \hat{k} \cdot \hat{n}$). Since $C_m$ varies with intensity [3] [4], our approach here, which assumes it is spatially constant, is not strictly valid. By the current scheme, however, the laser intensity is such that it is near the peak of the $C_m$ vs. intensity curve. Since the variation of $C_m$ in that region is modest, treating it as a constant serves to adequately demonstrate our results. Clearly, more sophisticated modeling, incorporating analytic fits to $C_m$ data, is required. It is also worth noting that, in the extreme case, obliquity could reduce the local intensity on target so as to inhibit plasma formation entirely, which suggests increasing the laser power to a few times that required to initiate plasma formation on a planar target, rather than near the peak of the $C_m$ vs. intensity curve.

It is possible for an irregularly shaped target to ablate in such a way as to create a torque about the center of mass, resulting in spin. Additionally, it is known that some space debris fragments are already spinning [2], which means that interactions with a laser may alter the spin frequency, or may alter the orientation of the spin vector. When spin in an asymmetric debris fragment is present (or is induced), the laser/target interaction will vary from shot to shot, resulting each time in a different impulse, which leaves a complex scenario for targeting and re-acquisition, as well as for the de-orbiting scheme in general. One might argue that spin is beneficial; with several hundred engagements, the range of possible orientations becomes well sampled, and the overall effect will tend toward the mean, producing results similar to those presented in Figure 4.

Finally, we mention the likelihood of structural modification during the course of a de-orbiting campaign. By momentum conservation, the fraction of ablated mass of a given target is approximately $\Delta v/v_A$. With $\Delta v \sim 100$ m s$^{-1}$, and the ablation flow speed $v_A \sim 1000$ m s$^{-1}$, this implies that about 10% of the mass will be ablated prior to atmospheric capture. The stepwise deformation of a target, and the evolution of the delivered impulse is probably best studied experimentally.

Continued exploration of these issues, coordinated with laboratory experiments, using a variety of materials, shapes, and laser pulse formats, will be essential to the process of downselecting from the range of possible laser configurations and acquisition/targeting strategies currently under consideration as candidates for laser debris remediation.

# ACKNOWLEDGMENTS

This work was performed under the auspices of the U.S. Department of Energy by Lawrence Livermore National Laboratory under Contract DE-AC52-07NA27344.